# An operational window for radiation-resistant materials based on sequentially healing grain interiors and boundaries

Xiangyan Li[1], Yichun Xu[1], C. S. Liu[1*], B. C. Pan[2], Yunfeng Liang[1,3], Q. F. Fang[1], Jun- Ling Chen[4], G.-N. Luo[4], Zhiguang Wang[5], Y. Dai[6]

[1]Key Laboratory of Materials Physics, Institute of Solid State Physics, Chinese Academy of Sciences, P. O. Box 1129, Hefei 230031, P. R. China.

[2]Hefei National Laboratory for Physical Sciences at Microscale and Department of Physics, University of Science and Technology of China, Hefei 230026, P. R. China.

[3]Environment and Resource System Engineering, Kyoto University, Kyoto 615-8540, Japan.

[4]Institute of Plasma Physics, Chinese Academy of Sciences, Hefei 230031, P. R. China.

[5]Institute of Modern Physics, Chinese Academy of Sciences, Lanzhou 730000, P. R. China.

[6]Spallation Neutron Source Division, Paul Scherrer Institut, 5252 Villigen PSI, Switzerland.

*To whom correspondence should be addressed. E-mail: csliu@issp.ac.cn

**Design of nuclear materials with high radiation-tolerance has great significance[1], especially for the next generation of nuclear energy systems[2,3]. Response of nano- and poly-crystals to irradiation depends on the radiation temperature, dose-rate and grain size[4-13]. However the dependencies had been studied and interpreted individually, and thus severely lacking is the ability to predict radiation performance of materials in extreme environments. Here we propose an operational window for radiation-resistant materials, which is based on a perspective of interactions among irradiation-induced interstitials, vacancies, and grain boundaries. Using atomic simulations, we find that healing grain boundaries needs much longer time than healing grain interiors. Not been noticed before, this finding suggests priority should be thereafter given to recovery of the grain boundary itself. This large disparity in healing time is reflected in the spectra of defects-recombination energy barriers by the presence of one high-barrier peak in addition to the peak of low barriers. The insight gained from the study instigates new avenues for examining the role of grain boundaries in healing the material. In particular, we sketch out the radiation-endurance window in the parameter space of temperature, dose-rate and grain size. The window helps evaluate material performance and develop resistant materials against radiation damage.**

Grain boundaries (GBs) or interfaces are prevalent in poly- and nano-crystalline materials, governing many of their properties including mass transport, mechanical strength, and deformation[14,15]. GBs are expected to be effective sinks for all types of defects[16–19], and generally considered as a detriment to material performance[20–23]. Recently, it has been proposed that advantage could be taken of GBs to absorb irradiation-generated defects like interstitials and vacancies, and then used to heal materials[1,19]. The positive role of GBs has been supported by some experiments[4–7] and simulations[18,19,24–28]. This is extremely beneficial to nuclear energy systems where highly radiation-tolerant materials are in demand[2,3].

Notwithstanding, overlooked is the fact that the GB has high energy and comes with an inherent instability, susceptible to radiation damage in GBs. This has been revealed by experiments[4–7]. Indeed GBs worsen radiation damage, depending on irradiation conditions (temperature and dose-rates) and grain sizes. In the case of gold, the nano-crystalline material is more radiation-tolerant at room temperature, while the poly-crystalline counterpart is more tolerant at extremely low temperatures ~15K[5]. Some nano-structured materials are stable to amorphization for a low dose-rate, while they undergo the crystalline-amorphous transformation when dose-rates increase[7]. Small grain-sized systems are more susceptible to irradiation-induced amorphization[8–11], contrary to the prediction [1,19]. These rather different irradiation responses show on one side that GBs can catalyze the recombination of interstitials and vacancies, but on the other side that GB can lose their role as sinks and catalysts. Therefore the critical issue is to understand why and how the GB plays positive or negative role during radiation. To address this issue, knowledge of damaging and healing the GB itself is essential, before understanding the mechanism of healing damage in materials via GBs by absorbing and removing defects.

To show how the system containing the GB is damaged and healed, we first performed MD simulations of collision cascades in a model tilt symmetric GB $\Sigma 5(310)/[001]$ in $\alpha$-Fe at 1000 K up to 4012 picoseconds (ps). The model geometry and the GB core structure are shown in Figs S1 and S2 in Supplementary Information. Several typical snapshots of defect generation and annihilation are shown in Fig. 1. The whole process is vividly visualized in the Movie S1 (see Supplementary Information). At about 0.5 ps, the cascade reaches its maximal size of about 3 nm and the cascade center is basically located at the GB plane (Fig. 1a). Interstitials move towards the GB, and after about 12 ps all the interstitials are trapped within the GB with a vacancy-dominated defect structure left in grain interiors (Fig. 1b). At about 267 ps, the two vacancies near the GB are completely removed, and consequently the grain interior is perfectly recovered (Fig. 1c). It seems that the whole system has been healed by this time, as observed previously[19].

However, when viewing along the GB, surprisingly we find that there still exists a large number of defects within the GB. After examining the vacancy and interstitial distribution (Fig. 1d-i), clearly observed is a deficient zone of atoms (DZ) within the GB. DZ consists of a large number of vacancies and has a size of about 3 nm with interstitials surrounding it uniformly. These defects within the GB are removed via interstitial diffusion into DZ or occasional vacancy emission from DZ (Movie S1 in Supplementary Information). Such process is evidently different from that in healing grain interiors[19]. By 4012 ps (Fig. 1i), only one Frenkel pair survives within the GB, and as a consequence both the grain interior and boundary are truly healed.

Apparently, our simulation strongly demonstrates that the damaged system heals via two processes. One is the healing of grain interiors from 0.5 to 267 ps, and the other is the healing of GBs up to 4012 ps. The difference in time scales implies distinctly different processes of defects evolution. As observed in Movie S1 (see Supplementary Information), the grain interior is healed via defect recombination and segregation, while the GB itself is healed via defect annihilation around DZ. Importantly, we find two characteristic times, $t_1$=267 ps and $t_2$=4012 ps from our simulations. Only if the time interval between two collision cascades $\tau_0$ (related to the dose-rate) is larger than $t_2$ is the whole system completely healed, which then enables the GB to absorb upcoming interstitials and vacancies in grain interiors sustainably.

To understand quantitatively the observed difference in healing of the grain interior and boundary, we calculated the activation energy spectrum[29]. The spectrum, suggestive of defect-recombination event distribution on energy barriers, can be derived from defect numbers evolution during irradiation (Fig. 2a), and used to describe the kinetics of the damaged system as a whole. As shown in Fig. 2a, the number of defect increases sharply from 0 to 1.0 ps, followed by an abrupt decay from 1.0 to 10 ps and a slow decay from 10 to 4012 ps. The big difference in the two decay rates suggests occurrence of events with different energy barriers. Interestingly, we consistently observed two Gaussian-like peaks with well-defined ranges of energy barriers in Fig. 2b. One is centered at 0.11 eV ($P_1$) and the other at 0.60 eV ($P_2$), which give rise to two decay rates of defect numbers in Fig. 2a.

The results indicate that the entire spectrum (Fig. 2b) originates from two different contributions of defect annihilation within the grain interior and boundary. In the partial spectra Fig. 2c-e both low energy and high-energy peaks appear, but only the low-energy peak is present in Fig. 2d and f, indicating that high temperatures are required to heal the GB. Figure 2d and f also indicate the preferential segregation of interstitials over vacancies into the GB. We also note that the small peak $P_9$ in Fig. 2f, not present in the total spectrum (Fig. 2b), arises from the GB absorption of vacancies nearby rather than annihilation.

On the basis of the results from simulations and support by the spectra, we established an overall perspective about interactions among GBs, irradiation-produced interstitials and vacancies (Fig. 3). The involved processes include segregation of interstitials and vacancies (processes *I* and *II*), their annihilation and respective diffusion in the bulk, near the GB and within the GB (processes *1*, *2* and *3* ). Processes *I*, *II* and *1* (noticed in the previous work[19]), and *3* (noticed in reference[30]) contribute to healing grain interiors. Process *2* is critical to healing the GB, which has never been noticed before.

In the next step, we calculated the energetic and kinetics of these processes using molecular statics (MS) and temperature-accelerated dynamics (TAD) methods[31,32], focusing on processes *I*, *II*, *1* and *2*. For processes *I* and *II*, our calculations reveal preferential absorption of interstitials by GBs because of larger segregation energy (2.7/0.5 eV), interaction range (14.8/9.4 Å) and lower diffusion barriers (0.33/0.63 eV) for interstitials/vacancies (Figs. S3-S5 in Supplementary Information), in agreement with previous results[19, 26, 27]. Further calculations indicate that, there exists level of GB width (interaction range) and GB depth (sink strength) for interstitials (15.6 Å, 0.78) and vacancies (11.6 Å, 0.36) in general to a certain extent (Figs. S6 and S7, Table S1 in Supplementary Information). Moreover, segregation energies for interstitials and vacancies both have strong correlations with GB energies (Fig. S8 in Supplementary Information). The GB depth needed for interstitials or vacancies to be absorbed by GBs depends on bulk defect concentration (Fig. S9, Tables S1 and S2 in Supplementary Information).

For process *1*, the interstitial has been located at the GB, resulting in the formation of a spontaneous annihilation region centered at the interstitial (Fig. S10-S12 in Supplementary

Information) and enhanced vacancy diffusion near the GB (Figs. S13 and S14 in Supplementary Information). Similarly, vacancies near the GB will modify the sink for interstitials compared with that in process *1*. If the vacancy is located within the above annihilation region (a limited range of about 10 Å), it will recombine with the interstitial, otherwise the interstitial has to overcome a high energy barrier of about 2.8 eV (Figs. S15 and S16 in Supplementary Information) to migrate out of the GB. The annihilation event involves a concerted motion of multiple atoms, with a distribution in both annihilation energy barriers and the number of atoms involved (Figs. S17-S19, Tables S3-S5, Movie S2 in Supplementary Information).

For process *2*, although interstitials and vacancies within the GB can source from the segregation processes *I* and *II*, but it should be stressed that interstitials and vacancies can also be more easily generated within the GB due to much lower defect formation energies there (Figs. S3 and S4 in Supplementary Information). As observed in Movie S1 (see Supplementary Information), irradiation-produced vacancies form a deficient zone of atoms (namely DZ) surrounded by interstitials within the GB. Careful examination shows that interstitials and vacancies are distributed at the bottom of the sink (Figs. S3 and S4 in Supplementary Information), where the defect formation energies reach a minimal value. Thus, the formation of DZ and distribution of interstitials within the GB have a close relation to the defect formation energy profile at the GB. This implies that our observed phenomenon during irradiation is popular because of similar features on defect formation energy profiles in most GBs (Fig. S6 in Supplementary Information).

When the interstitial is very close to DZ, the annihilation is similar to that in process *1* (Figs. S17-S19, Tables S3-S5, Movie S3 in Supplementary Information). For interstitial far apart from DZ, it has to diffuse individually before encountering DZ. Our TAD calculations indicate that interstitials diffuse concertedly within the GB involving three atoms with displacements larger than 0.6 Å and ten other atoms with smaller displacements (Figs. S20 and S21, Movie S4 in Supplementary Information). The easiest transition is parallel to the tilt axis which still has a barrier of 0.59 eV. Vacancy diffusion also is a little difficult within the GB (Fig. S20 in Supplementary Information).

Combining atomic calculations of processes in Fig. 3 with the spectra in Fig. 2b-f, we know that the low-energy peak in Fig. 2b stems from annihilation of close Frenkel pairs, while both segregation and annihilation contribute to the peaks of low energy barriers in Fig. 2c-f. The high-energy peak arises from the formation of the DZ and difficult diffusion of interstitials within the GB, which naturally make the probability smaller of interstitials and vacancies encountering each other within the GB. Now the spectra can shed light on understanding the large difference in healing of the grain interior and boundary. Based on the activation energy and its dependence on the system temperature, we can accordingly judge whether defects can survive in different regions. At extremely low temperatures, a large portion of defects with barriers about 0.1 eV (Fig. 2b-f) are practically immobile and will survive within both the grain interior and GB. These defects will therefore accumulate with time constantly. As the temperature increases defects will recombine or segregate into the GB. The defects, corresponding to the low temperature peaks, begin to lose their role in defect accumulation. Only those defects with energy barriers about 0.6 eV (Fig. 2b) can survive within the GB. When the temperature is high enough to anneal defects out within the GB, the GB can maintain to be a perfect sink. In other words, the spectra mainly indicate a sequential healing of the grain interior and the GB itself. Meanwhile, a stepwise temperature-dependent irradiation response of materials can be expected, in agreement with experimental results[5].

On the basis of the overall perspective mentioned above, we propose a operational window for nuclear materials in the parameter space of temperatures, radiation dose-rates and grain sizes (Fig.4 and Fig. S23 in Supplementary Information). In details, the region between two grains are divided into the annihilation region, the low barrier region, the buffer region, and the bulk region (Fig. 4a), where defects exhibit obviously different behaviors in defect annihilation or diffusion. To determine the range of buffer region ($L$) and its dependence on temperature, Fig. 4b presents the distance that an interstitial or a vacancy diffuse within a given time interval between two cascades at a certain temperature. According to irradiation responses, the parameter space of grain sizes and temperatures can be divided into three functional regimes.

In the regime A, both the interstitial and vacancy cannot migrate into the GB because the grain size is larger than the distance that the defects can diffuse within the given time interval. If the system temperature is extremely low (lower than 40 K in Fig. 2b), all of the atomic processes are inactive and defects will accumulate in the whole system (as experimentally observed results in the system of Au with a size of 23 nm at 15 K[5]). As the temperature increases, defects close to each other can recombine and only well-separated interstitials and vacancies can survive (Fig. 2b). If the temperature is high enough to activate interstitials (Fig. S14 in Supplementary Information) the system can recover well (as experimentally observed results in the system of Cu with a size of 250 nm at 77 K[10]). In the regime B, the grain size is so small that interstitials can diffuse into the GB with vacancies left in grain interiors. The system will exhibit inferior radiation performance (as experimentally observed results in the system of FeAl with a size of 160 nm at room temperature[7]). In the regime C, both interstitials and vacancies can migrate into the GB and the GB works well in serving as sinks for defects. Hence the system shows the best radiation damage tolerance (as experimentally observed results in the system of FeAl with a size of 35 nm[7] and that of Au with a size of 23 nm at room temperature [5]).

Additionally, as indicated by the black line in Fig. 4b, there should exists a critical size below which the GB will lose efficiency as sinks because of difficulty in healing the GB. From the consideration of the formation of the DZ induced by PKA with 3 keV energy in our simulations, and the low mobility of interstitials within the GB plane, the critical size is about 10 nm. The precise value of the critical size should be related to PKA energies. Furthermore, the difficulty in healing the GB predicts that small systems are more susceptible to damage (as experimentally observed results in the system of Cu with a size of 2.8 nm at 76 K[10]), contrary to the previous prediction [1, 19]. Another factor is the radiation dose-rates, which are related to defect migration time (Fig. S23 in Supplementary Information).

Based on above calculations and experimental results, we believe that there exists a window where materials exhibit high radiation tolerance, valuable to the design and utilization of highly radiation-resistant materials. Meanwhile, there exist regimes where GBs lose their positive roles as sinks and catalysts. To recover radiation-induced damage perfectly, it is essential to heal the GB sink itself, besides healing grain interiors via the sink absorbing defects. This concept can be applicable to other types of boundaries like interfaces of interphase because of easily damaged feature of interfaces. In addition, slow processes like GB healing should be incorporated into predictive models of material lifetimes.

**Methods**

**GB structure modeling**. The grain boundary (GB) used in this work is a $\Sigma 5$ (310)/[001] symmetric tilt GB ($\Sigma$ indicates the degree of geometrical coincidence at a GB) in bcc iron containing 95,520 atoms with a size of about $180 \times 70 \times 90$ Å$^3$. Periodic boundary conditions were applied in the two directions parallel to the GB plane, but the fixed boundary condition is in the direction normal to the GB plane. The simulation cells consist of a moving region sandwiched between two rigid regions (see Supplementary Information Fig. S1). The GB energy is minimized through the relaxation of all non-rigid atoms and the rigid-body translations of one grain relative to the other in all three Cartesian directions at 0 K. The lowest-energy GB structure has energy of 0.99 J/m$^2$. And a smaller system having 3792 atoms with a size of about $70 \times 30 \times 20$ Å$^3$ is used for the calculation of vacancy (interstitial) formation energies, migration barriers and their annihilation barriers near the GB and within the GB in considering the computational time expense of temperature-accelerated dynamics (TAD)[31,32].

**Detains of MD.** Molecular dynamics (MD) was performed to study cascade-induced damage near the GB at 1000 K using the velocity-Verlet method for the numerical integration, and the simulations ran for about 4 nano-seconds. An atom at 23 Å on one side of the GB and located at the center of the plane parallel to the GB plane was selected as the PKA. The atom was given 3 keV of kinetic energy with its velocity directed perpendicularly toward the GB. The atoms in the outermost layers of the moving region, with a thickness of two times the lattice constant, were coupled with a velocity-rescaling thermostat to absorb the cascade energy and maintain the system temperature at 1000 K. The embedded-atom-method (EAM) interatomic potential developed by Mendelev *et al.*[33] was used to model the interatomic interaction, and the short-range form of the potential was splined to reproduce the high-energy empirical potential of Ziegler *et al*[34].

**Characterization of the damaged GB structure.** In the visualization of defect configurations, atoms were color coded to differetiate their potential energy. Atoms with energy deviation from the bulk value less than 0.1 eV were treated as non-defective and not shown. A vacancy is characterized as a 14-atom cluster consisting of its first and second nearest neighbors. And the cluster comprises of atoms less than 14 within the GB for a smaller coordination number. To count defect numbers during the radiation, the resulting interstitials and vacancies were identified using the Wigner-Seitz cell method[35]. The spectrum of activation energies[29] was adopted as an approach to kinetics of the damaged GB structure with a highly disordered region localized in space.

**Analysis of the activation energy spectrum.** Molecular statics calculations were carried out for vacancy and interstitial formation energies and segregation energies, while defect migration barriers were determined using the nudged elastic band (NEB) method[36]. The interstitial diffusion within the GB and the annihilation events near the GB and within the GB were investigated using TAD, because how these events proceed is non-intuitive. In the TAD calculation, the lower and high temperatures were set to 300 K and 1000 K, respectively based on the requirements that the method could monitor events occurring on a time scale of nano-seconds at the chosen high temperature. The minimum preexponential factor $\nu_{min}$ and an uncertainty $\delta$ were set at $\nu_{min} = 5 \times 10^{11}$ and $\delta = 0.05$.

**References and Notes**

1. Ackland, G. Controlling radiation damage. *Science* **327**, 1587-1588 (2010).

2. Samaras, M., Victoria, M. & Hoffelner, W. Nuclear energy materials prediction: aplication of the multi-scale modelling paradigm. *Nucl. Eng. Technol*. **41**, 1-10 (2009).

3. Hoffelner, W. Damage assessment in structural metallic materials for advanced nuclear plants. *J. Mater. Sci*. **45**, 2247-2257 (2010).

4. Rose, M., Balogh, A. G. & Hahn, H. Instability of irradiated induced defects in nanostructured materials. *Nucl. Instr. and Meth. B* **127**, 119-122 (1997).

5. Chimi, Y. *et al*. Accumulation and recovery of defects in ion-irraddiated nanocrystalline gold. *J. Nucl. Mater*. **297**, 355-357 (2001).

6. Kilmametov, A. R., Gunderrov, D. V., Valiev, R. Z., Balogh, A. G. & Hahn, H. Enhanced ion irradiation resistance of bulk nanocrystalline TiNi alloy. *Scr. Mater*. **59**, 1027-1030 (2008).

7. Kilmametov, A. et al. Radiaiton effects in bulk nanocrystalline FeAl alloy. *Radiat Eff. Defects Solids*. **167-8**, 631-639 (2012).

8. Kachurin, G. A. *et al*. Light particle irradiation effects in Si nanocrystals. *Nucl. Instr and Meth. B* **147**, 356-360 (1999).

9. Ridgway, M. C. *et al*. Ion-irradiation-induced preferential amorphization of Ge nanocrystals in silica. *Phys. Rev. B* **74**, 094107 (2005).

10. Johannessen, B. *et al*. Amorphization of embedded Cu nanocrystals by ion irradiation. *Appl. Phys. Lett*. **90**, 073119 (2007).

11. Jiang, W., Wang, H., Kim, I., Zhang, Y. & Weber, W. J. Amorphization of nanocrystalline

3C-SiC irradiated with Si+ ions. *J. Mater. Res*. **25**, 2341-2348 (2010).

12. Ovid′ko, I. A. & Sheinerman, A. G. Irradiation-induced amorphization process in nanocrystalline solids. *Appl. Phys. A* **81**, 1083-1088 (2005).

13. McClintock, D. A., Hoelzer, D. T., Sokolov, M. A. & Nanstad, R. K. Mechanical properties of neutron irradiated nanostructured ferritic alloy 14YWT. *J. Nucl. Mater*. **386-388**, 307-311 (2009).

14. Chen, M. *et al*. Deformation twinning in nanocrystalline aluminum. *Science* **300**, 1275-1277 (2003).

15. Tuller, H. L. Ionic conduction in nanocrystalline materials. *Solid State Ionics* **131**, 143 (2000).

16. Lejcek, P. *Grain Boundary Segregation in Metals* (Springer, Verlag, 2010).

17. Kingery, W. D. Plausible concepts necessary and sufficient for interpretation of ceramic grainboundary phenomena: I, grain-boundary characteristics, structure, and electrostatic potential. *J. Am. Ceram. Soc*. **57**, 1-8 (1974).

18. Samaras, M., Derlet, P. M., Van Swygenhoven, H. & Victoria, M. Computer Simulation of Displacement Cascades in Nanocrystalline Ni. *Phys. Rev. Lett*. **88**, 125505 (2002).

19. Bai, X. M., Voter, A. F., Hoagland, R. G., Nastasi, M. & Uberuaga, B. P. Efficient Annealing of Radiation Damage Near Grain Boundaries via Interstitial Emission. *Science* **327**, 1631-1634 (2010).

20. Yan, Y. *et al*. Impurity-Induced Structural Transformation of a MgO Grain Boundary. *Phys. Rev. Lett* **81**, 3675-3678 (1998).

21. Rainer, S., Anthony, T. P., Michael, W. F. Bismuth embrittlement of copper is an atomic size

effect. *Nature* **432**, 1008-1011 (2004).

22. Masatake, Y. *et al*. Grain Boundary Decohesion by Impurity Segregation in a Nickel-Sulfur System. *Science* **307**, 393-397 (2005).

23. Wang, Z. H. *et al*. Atom-resolved imaging of ordered defect superstructures at individual grain boundaries. *Nature* **479**, 380-383 (2011).

24. Samaras, M., Derlet, P. M. & Van Swygenhoven, H. Movement of interstitial clusters in stress gradients of grain boundaries. *Phys. Rev. B* **68**, 224111 (2003).

25. Samaras, M., Derlet, P. M., Van Swygenhoven, H. & Victoria, M. Atomic scale modelling of the primary damage state of irradiated fcc and bcc nanocrystalline metals. *J. Nucl. Mater*. **351**, 47-55 (2006).

26. Tschopp, M. A., Horstemeyer, M. F., Gao, F., Sun, X. & Khaleel, M. Energetic driving force for preferential binding of self-interstitial atoms to Fe grain boundaries over vacancies. *Scr. Mater*. **64**, 908-911 (2011).

27. Tschopp, M. A. *et al*. Probing grain boundary sink strength at the nanoscale: Energetics and length scales of vacancy and interstitial absorption by grain boundaries in *α*-Fe. *Phys. Rev. B* **85**, 064108 (2012).

28. Bai, X. M. *et al*. Role of atomic structure on grain boundary-defect interactions in Cu. *Phys. Rev. B* **85**, 214103 (2012).

29. Gibbs, M. R. J., Evetts, J. E. & Leake, J. A. Activation energy spectra and relaxation in amorphous materials. *J. Mater. Sci*. **18**, 278-288 (1983).

30. Gibson, J. B., Goland, A. N., Milgram, M. & Vineyard, G. H. Dynamics of radiation damage. *Phys. Rev*. **120**, 1229-1253 (1960).

31. Sørensen, M. R. & Voter, A. F. Temperature-accelerated dynamics for simulation of infrequent events. *J. Chem. Phys*. **112**, 9599-9606 (2000).

32. Voter, A. F., Montalenti, F. & Germann, T. C. Extending the time scale in atomistic simulation of materials. *Annu. Rev. Mater. Res*. **32**, 312-346 (2002).

33. Mendelev, M. I. *et al*. Development of new interatomic potentials appropriate for crystalline and liquid iron. *Philos. Mag*. **83**, 3977-3994 (2003).

34. Ziegler, J. Biersack, J. P. & Littmark, U. *The Stopping and Range of Ions in Solids* (Pergamon Press, New York, 1985).

35. Honeycutt, J. D. & Andersen, H. C. Molecular dynamics study of melting and Freezing of Small Lennard-Jones Clusters. *J. Phys. Chem*. **91**, 4950-4963 (1987).

36. Henkelman, G. & J′onsson, H. Improved tangent estimate in the nudged elastic band method for finding minimum energy paths and saddle points. *J. Chem. Phys*. **113**, 9978-9985 (2000).

37. Ziegler, J. F., Ziegler, M. D. & Biersack, J. P. SRIM C The stopping and range of ions in matter. *Nucl. Instrum. Methods Phys. Res. B* 268, 1818C1823 (2010).

**Acknowledgments** We thank G. J. Ackland, Guowen Meng, Changhui Ye, Wei Hua Wang, and X. Y. Qin for comments, help, and suggestions; X. B. Wu, and Xiang-Shan Kong for improvement of graphics. This work was supported by the National Magnetic Confinement Fusion Program (Grant No.: 2011GB108004), the Strategic Priority Research Program of Chinese Academy of Sciences (Grant No.: XDA03010303), the National Natural Science Foundation of China (No.:91026002), and by the Center for Computation Science, Hefei Institutes of Physical

Sciences.

**Author Contributions** C. S. L. discussed the results and directed the entire study. Most of the calculations were performed by X. L and Y. X. All authors contributed to the analysis and discussion of the data and the writing of the manuscript.

**Author Information** The authors declare no competing financial interests. Supplementary information accompanies this paper on www.nature.com/naturematerials. Reprints and permissions information is available online at http://www.nature.com/reprints. Correspondence and requests for materials should be addressed to C. S. L. (csliu@issp.ac.cn).

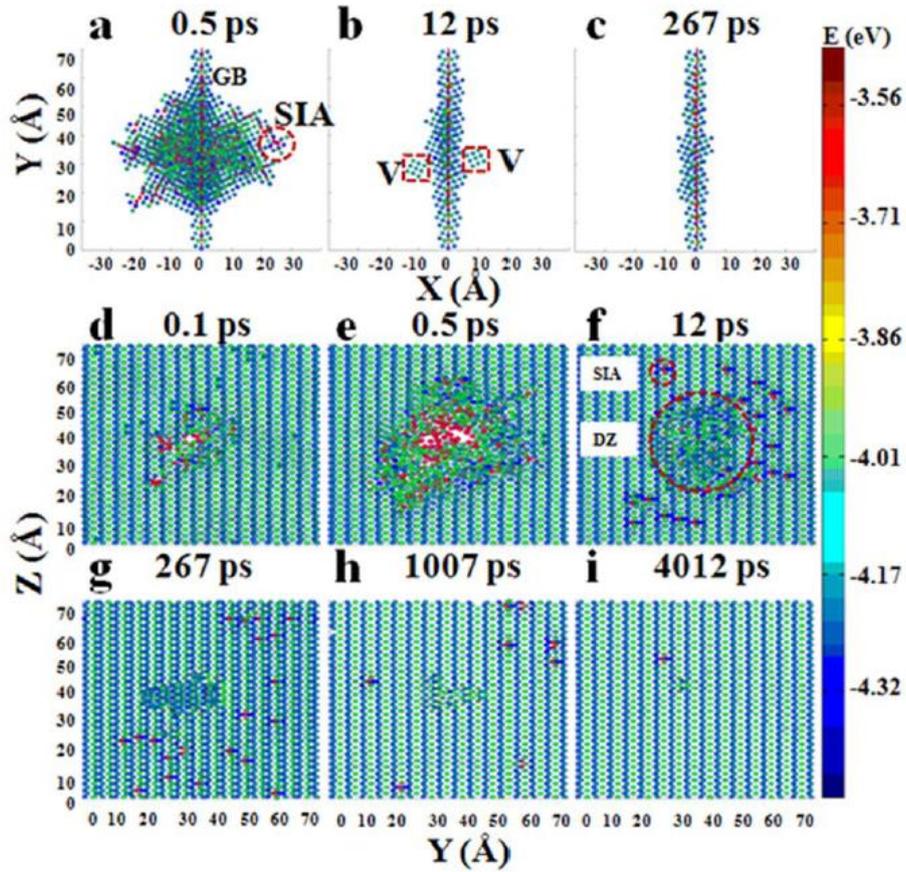

**Figure 1** | **The defect evolution within the grain interior and GB. a-c**, Defect evolution in grain interiors. **d-i**, Defect evolution within the GB. Here *V* is indicative of a vacancy while *DZ* is for deficient zone of atoms and *SIA* is for interstitials. A vacancy is characterized as a 14-atom cluster consisting of its first and second nearest neighbors. And the cluster comprises atoms less than 14 at the GB for a smaller coordination number than that in the bulk. Atoms are colored with their potential energy; atoms with energy deviation from the bulk value less than 0.1 eV are treated as non-defective and are not shown. X, Y and Z are along [310], [$\bar{1}$30] and [001], respectively.

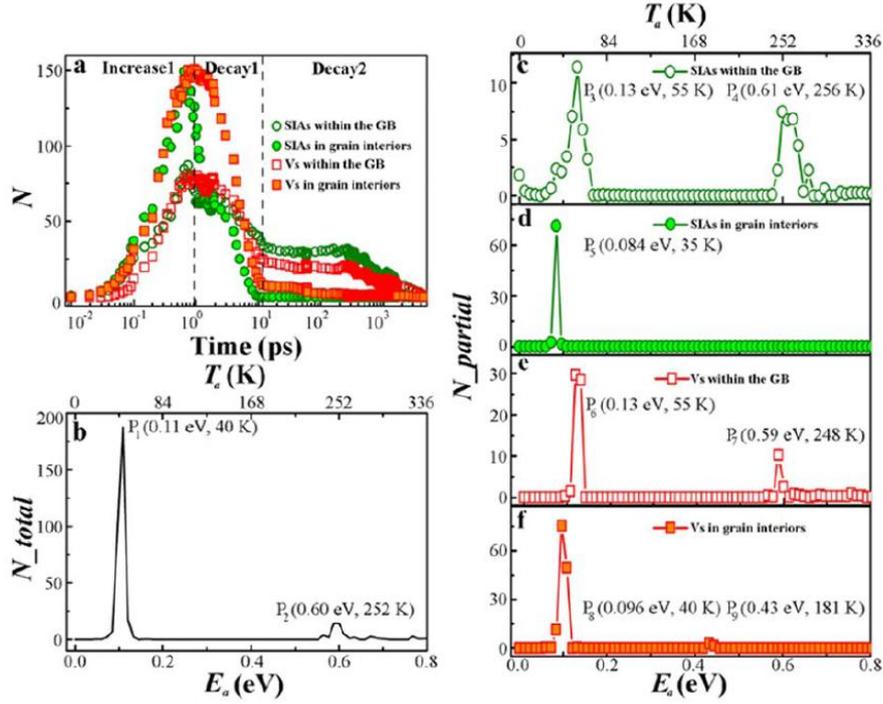

**Figure 2 | Variations of defect numbers as a function of time and activation energy spectra of the damaged system. a**, Evolution of vacancy and interstitial numbers $N$ with time within the grain interior and GB. Defect numbers increase sharply with time, followed by an abrupt decay and a slow decay (Increase1, Decay1 and Decay2 indicates three stages, respectively). **b**, The total activation energy spectrum derived from variations of total defect numbers with time. $N\_total$ is the total defect numbers. **c-f**, The partial spectra of activation energies. $N\_partial$ is the defect number of interstitials or vacancies. Assuming first-order kinetics holds for segregation and annihilation of radiation-produced defects, the defect number $N(t) = \sum_{i=1}^{n_1} e^{-t/t_i} = \sum_{j=1}^{n_2} f_j e^{-t/t_j}$ and the lifetime $t_j$ can be related to the activation energy $E_{aj}$ via $E_{aj} = k_B T ln(t_j \upsilon_0)$ during relaxation of the damaged structure. The distribution of $f_j$ in $E_{aj}$ is the activation energy spectrum. Activation temperature $T_a$ is obtained via harmonic approximation $t = 1/\upsilon_0 e^{E_a/k_B T_a}$ where $\upsilon_0$ is chosen as $10^{12}$/s and $k_B$ is the Boltzmann constant of $8.617 \times 10^{-5}$ eV/K; $E_a$ is activation energy. And $t$ is given a value of one second here, which is the time interval between

two successive cascades. The GB region is the bottom of the sink that GBs function as for defects. And the width of the region is about 5/10 Å for vacancies/interstitials (see Supplementary Information, Figs S3 and S4).

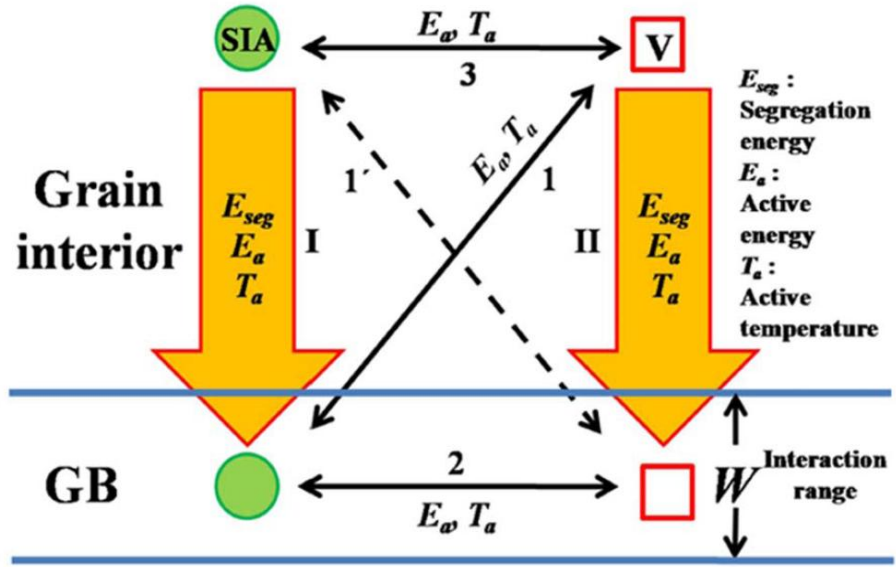

**Figure 3 | Illustration of interactions among vacancies, interstitials and GBs.** Here green spheres are indicative of interstitials and red squares are for vacancies. Processes *I* and *II* are segregation of interstitials and vacancies characterized by segregation energy $E_{seg}$, activation energy $E_a$ and activation temperature $T_a$. $E_{seg}$ is reduction of defect formation energies compared with that in the bulk. Process *1*, *2*, and *3* are annihilation of close Frenkel pairs or respective diffusions of interstitials and vacancies as they are far apart, in the bulk, near the GB, and within the GB. *W* is the width of interaction range (Figs S3 and S4 in Supplementary Information). The process *1* has been described as interstitial emission[19]. The process *1'* mainly exists in some special systems where vacancies have higher mobility than interstitials or occurs in annihilation of interstitials near the GB and vacancies within the DZ. Processes *I*, *II*, *1*, and *3* heal the grain interior while annihilation in the process *2* contributes to healing the GB.

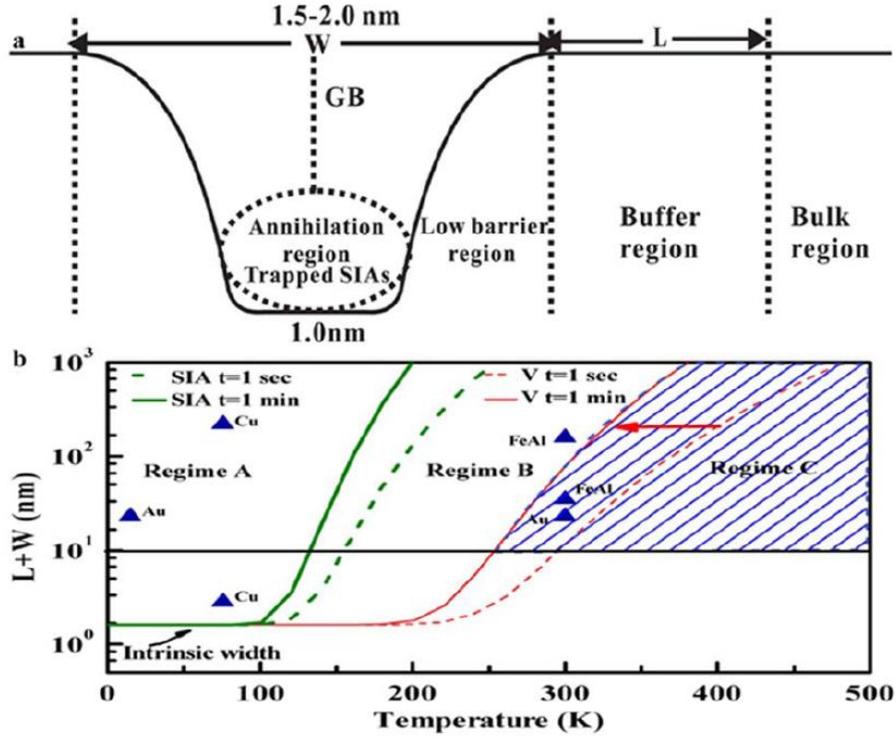

**Figure 4 | GB functional regions and the estimated optimal grain size based on interstitial and vacancy behaviors within the grain interior and GB.** Here $W$ is the interaction width between the GB and defects (Fig. S6 in Supplementary Information). **a**, Division of the space between two grains considering vacancy and interstitial formation energies and migration energies (see Supplementary Information, Figs S3-S5). The annihilation region, consisting of unstable sites (see Supplementary Information, Figs S10 and S17), surrounds the interstitial trapped at the GB. The low barrier region is GB-enhanced defect diffusion region. The buffer region with a width of $L$ is directional defect-diffusion region driven by the defect concentration gradient between the low barrier region and the bulk region. The bulk region has no biased driving force for defect diffusion and the process *3* (Fig. 3) works there. **b**, Different regimes in the parameter space of the grain size and temperature. $L = \sqrt{6Dt}$ and the diffusion coefficient $D$ for a single vacancy or interstitial at temperature $T$ is given by $D \approx d^2 \upsilon_0 exp(-E_a / k_B T)$. $d$ is the jump distance, which

is $\frac{\sqrt{3}}{2}a$ for a bcc lattice and $a$ is the lattice constant. As a conservative estimation, activation energy $E_a$ in the bulk is used having a value 0.63 eV for vacancies and 0.33 eV for interstitials. $t$ is the time interval between collision cascades given two values one second and minute, which relates to dose-rates. For a low dose-rate a large time interval, the border of the regime C extends towards left as the red arrow indicates. SRIM[37] gives several dpa per year for radiation of PKA with 3 keV to $\alpha$-Fe of 5 nm, given cascade time interval one second. To make comparison with available experimental results, some samples with a certain grain size and irradiated at a certain temperature are marked using a blue filled triangle.